\documentclass[12pt]{iopart}
\pdfoutput=1
\usepackage{graphicx}
\begin{document}

\title[An extensive comparison of anisotropies in MBE grown (Ga,Mn)As material.]{An extensive comparison of anisotropies in MBE grown (Ga,Mn)As material.}

\author{C Gould$^1$, S Mark$^1$, K Pappert$^1$, G Dengel$^1$, J Wenisch$^1$, R P Campion$^2$, A W Rushforth $^2$ D Chiba$^{4,5}$, Z Li$^3$, X Liu$^6$, W Van Roy$^3$, H Ohno$^{5,4}$,  J K Furdyna$^6$, B Gallagher$^2$, K Brunner$^1$, G Schmidt$^1$, L W Molenkamp$^1$}

\address{$^1$Physikalisches Institut (EP3), Universit\"{a}t W\"{u}rzburg, Am
Hubland, D-97074 W\"{u}rzburg, Germany}

\address{$^2$School of Physics and Astronomy, University of Nottingham, Nottingham NG7 2RD, United Kingdom}

\address{$^3$IMEC, Kapeldreef 75, B-3001 Leuven, Belgium}

\address{$^4$Semiconductor Spintronics Project, Exploratory Research for Advanced Technology, Japan Science and Technology Agency, Kitamemachi 1-18, Aoba-ku, Sendai 980-0023, Japan}

\address{$^5$Laboratory for Nanoelectronics and Spintronics,
Research Institute of Electrical Communication, Tohoku University,
Katahira 2-1-1, Aoba-ku, Sendai 980-8577, Japan}

\address{$^6$Department of Physics, University of Notre Dame, Notre Dame, Indiana 46556, USA}

\ead{gould@physik.uni-wuerzburg.de}

\begin{abstract}

This paper reports on a detailed magnetotransport investigation of
the magnetic anisotropies of (Ga,Mn)As layers produced by various
sources worldwide. Using anisotropy fingerprints to identify
contributions of the various higher order anisotropy terms, we show
that the presence of both a [100] and a [110] uniaxial anisotropy in
addition to the primary ([100] + [010]) anisotropy is common to all
medium doped (Ga,Mn)As layers typically used in transport
measurement, with the amplitude of these uniaxial terms being
characteristic of the individual layers.

\end{abstract}

\maketitle


\section{Introduction}\label{sec:intro}

A key prototypical material for investigations into spintronics is
the ferromagnetic semiconductor (Ga,Mn)As. The marriage of magnetic
and semiconductor properties brought about by strong spin-orbit
coupling, which ties the density of states of this material to its
magnetic properties, offers a host of new and exploitable
magnetotransport effects. As investigations into this material
continue to progress, it has become clear that a detailed
understanding of the underlying magnetic anisotropy is a key issue
in device design and optimization.

This magnetic anisotropy in (Ga,Mn)As is very rich and complicated,
which led to various reports over the past decade which
superficially appeared to be contradictory. In general, depending on
growth strain, doping density and temperature, the material can have
an easy axis of magnetization either perpendicular to plane, or in
the layer plane \cite{PhysicaE,Liu2003}, and in the latter case, the
primary anisotropy can be either biaxial along [100] and [010], or
uniaxial along either [110] or $[\bar{1}10]$
\cite{SawickiAnis,Sawicki_InPlane}.

It is generally accepted that for typical medium doped transport
samples with $\sim$ 3 to 6 \% Mn, grown with compressive strain and
measured at 4 K, the primary anisotropy is a biaxial term with easy
axes along the [100] and [010] crystal directions. Second order
terms are also widely reported in the form of a uniaxial easy axis
along [110] or $[\bar{1}10]$ \cite{Roukes} of various relative
strengths, or a uniaxial along [010] \cite{TAMR1}.

Since it is well known in the community that the detailed properties
of (Ga,Mn)As depend on exact growth conditions such as substrate
temperature, growth rate, flux ratios, etc. \cite{FurdynaGrowth,
CampionGrowth, Giselagrowth}, it was initially widely assumed that
the observation of these different higher order anisotropy terms
were primarily a result of the distinct properties of the various
layers used.

It is now realized that part of the confusion arose from the fact
that the various transport measurement from which these terms had
been extracted differ in sensitivity to the different anisotropy
terms. For example, in the non-volatile Tunneling Anisotropic
Magneoresistance (TAMR) experiments \cite{TAMR1}, the [010] plays a
crucial role whereas the [110] anisotropy is nearly irrelevant
because it only has significant impact for volatile effects
occurring at higher fields. On the other hand, the [010] term plays
only a secondary role in the planar Hall measurements of hall bars
along the [110] direction \cite{Roukes}.

Moreover, because these second order uniaxial anisotropy terms are
significantly weaker than the primary biaxial anisotropy, they
cannot be reliably characterized by direct magnetization
measurements such as SQUID (Superconducting Quantum Interference
Device) or VSM (Vibrating Sample Magnetometer). The challenge of
fully characterizing the complex anisotropies in (Ga,Mn)As was
recently successfully addressed with the development of an
"anisotropy fingerprint" technique \cite{Fingerprints} which
consists of taking magnetotransport measurements for magnetic fields
swept in multiple directions.

Using this method, we recently investigated \cite{NJPfingerprints}
various transport samples produced on wafers grown in a given
Molecular Beam Epitaxy (MBE) system at W\"{u}rzburg University, and
showed that in all cases, a detailed investigation revealed the
presence of both [010] and [110] uniaxial terms, with the sign and
relative amplitude of these two terms varying from sample to sample.
In the present paper, we expand this investigation to samples grown
by multiple groups and show that indeed the co-existence of all
anisotropy terms is a general property of (Ga,Mn)As, and that only
the relative strength of the terms is characteristic of the layer
growth.

\section{Anisotropy fingerprints}\label{sec:aniso}

\begin{figure} [h!] \centering
\includegraphics[width=0.85\textwidth]{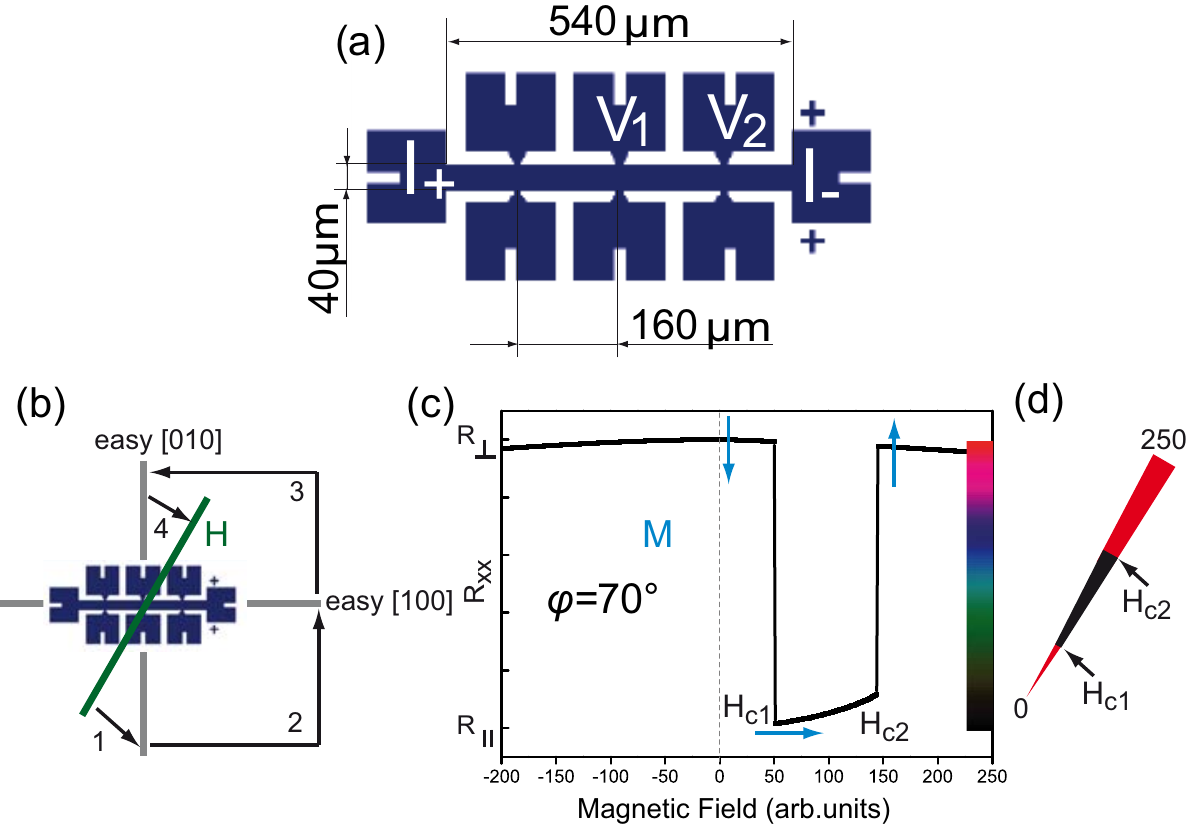}
\caption{\label{Hallbar} \textit{a: Layout of the Hall bar used in
the experiments. b: Configuration for the simulation of a
magnetoresistance scan along $\phi = 70^{\circ}$ (c) showing the two
switching event $H_{c1}$ and $H_{c2}$ corresponding to the two
subsequent $90^{\circ}$ domain wall propagation events. This data is
then converted (d) to a sector of a resistance polar plot.}}
\end{figure}

All investigations are performed using Hall bars of the
configuration shown in Fig.~\ref{Hallbar}a produced by standard
optical lithography followed by chemically assisted ion beam etching
(CAIBE). Magnetoresistance measurements are carried out in a
magnetocryostat equipped with a vector field magnet capable of
producing fields of up to 300 mT in any spatial direction. For the
measurement discussed in this paper, fields are always applied in
the plane of the sample, and the direction of the magnetic field is
given by the angle $\phi$ relative to the [100] crystal direction.

(Ga,Mn)As exhibits a strongly anisotropic magnetoresistance (AMR)
where the resistivity $\rho_{\perp}$ for current flowing
perpendicular to the direction of magnetization is larger than
$\rho_{\parallel}$ for current along the magnetization
\cite{Baxter}. As a result of this anisotropy in the resistivity
tensor, the longitudinal resistivity $\rho_{xx}$ is given by
\cite{Jan1957,McGuire}:

\begin{center}
{\parbox{15cm}{
\begin{equation}
\label{eqn:Rxx}
\rho_{xx}=\rho_{\bot}-(\rho_{\bot}-\rho_{||})\cos^2(\vartheta),
\end{equation}
}}
\end{center}

where $\vartheta$ is the angle between the direction of
magnetization and the current. Note that there is also a dependence
of the resistivity on the angle between the direction of
magnetization and the underlying crystal orientation
\cite{Rushforth}. This additional term modifies the resistivity
value for a given magnetization directions, but does not effect the
field position of the magnetization reorientation events, and can
thus be neglected for the purposes of the present analysis.

\begin{figure}[h!] \centering
\includegraphics[width=.4\textwidth]{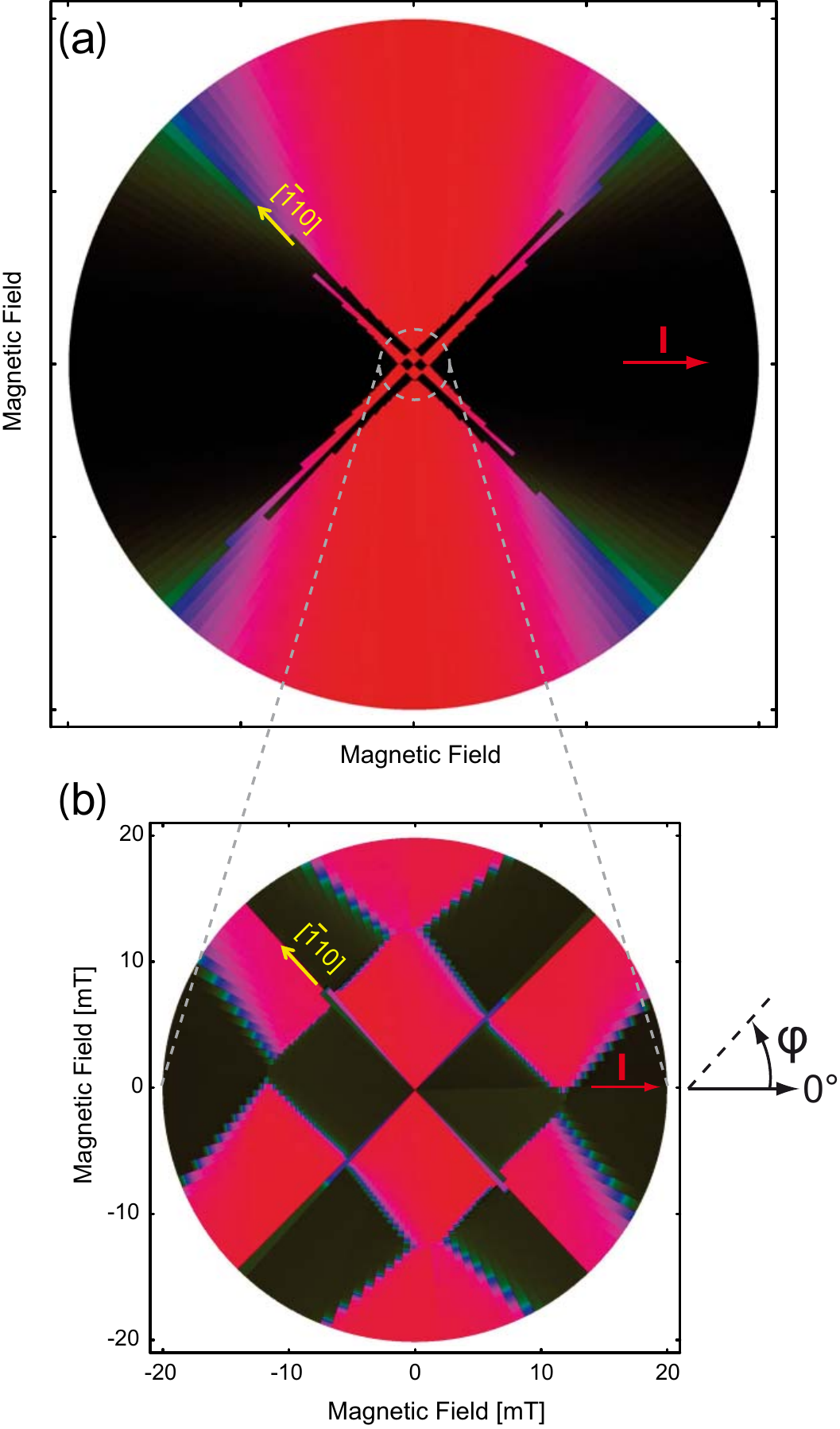}
\caption{\label{Zoom} \textit{a: Simulation of a full resistance
polar plot comprised of sectors as in Fig. \ref{Hallbar}. b:
measurement of the inner region of the polar plot. The red I
indicates the direction of current flow during the measurement.}}
\end{figure}

For each sample, we measure the four terminal longitudinal
resistance using the lead configuration given in Fig. \ref{Hallbar}a
by passing a current from the $I_+$ to the $I_-$ contacts, and
measuring the voltage between $V_1$ and $V_2$. We scan the magnetic
field from -300 mT to +300 mT along a given direction $\phi$, and
repeat this procedure for multiple angles. A simulation of such a
scan for the case of $\phi = 70^{\circ}$ is given in Fig.
\ref{Hallbar}c, and shows two switching events, labeled $H_{c1}$ and
$H_{c2}$ associated with the two sequential $90^{\circ}$ domain wall
nucleation/propagation events which account for the magnetization
reversal in this material \cite{Welp03PRL}. In order to analyze the
data, the positive field half of each of these scans are converted
to a sector of a polar plot as shown in Fig. \ref{Hallbar}d. The two
switching events then show up as abrupt color changes as indicated
in the figure. The compilation of all the sectors required for a
full revolution produces an anisotropy fingerprint resistance polar
plot as the one simulated in Fig.~\ref{Zoom}a.

For the purposes of characterizing the various anisotropy terms, the
most important part of the data is the innermost region whose
boundaries are formed by the loci of first switching events
($H_{c1}$). Fig.~\ref{Zoom}b shows a zoomed in view of this region
for an experimental measurement on a characteristic piece of
(Ga,Mn)As.

\begin{figure}[h!] \centering
\includegraphics[width=0.85\textwidth]{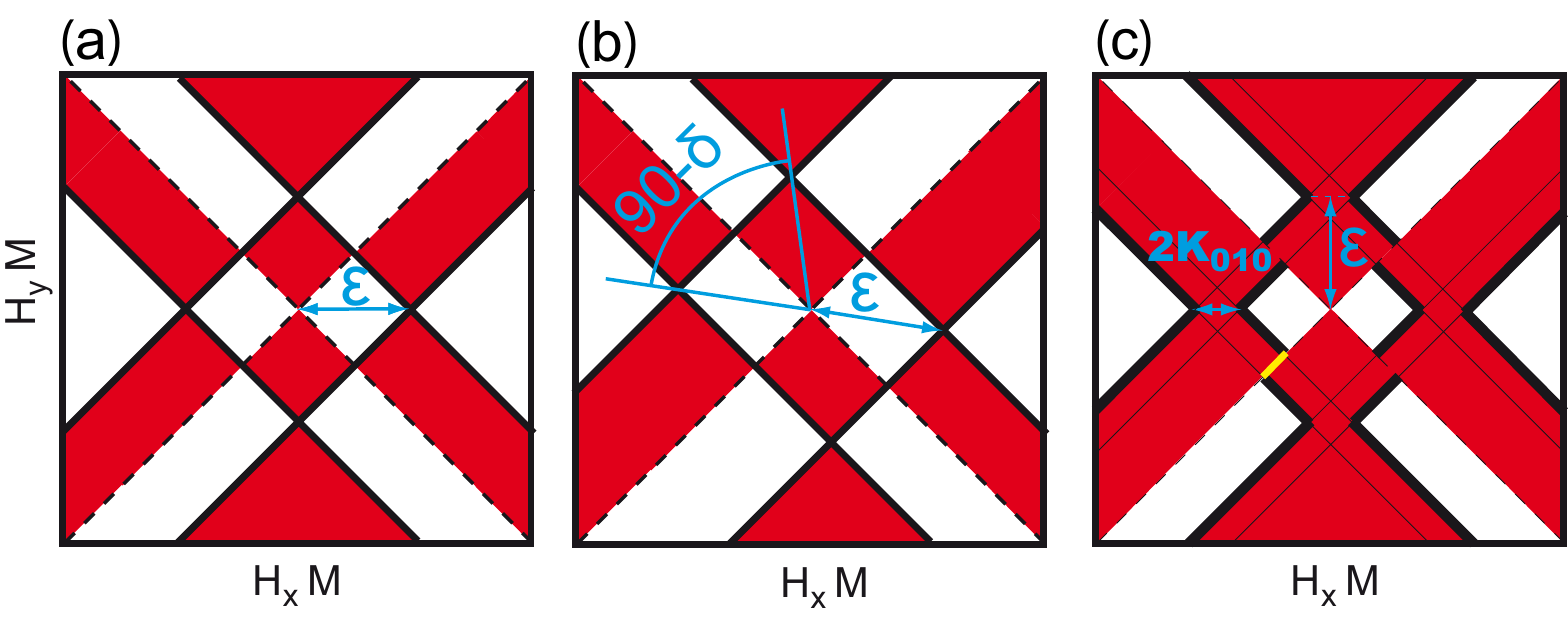}
\caption{\label{squares} \textit{Sketches of the expected shape of
the inner region for a) a samples with only a ([100] and [010])
biaxial anisotropy. b) a sample with a biaxial plus a $[\bar{1}10]$
uniaxial easy axis, and c) a samples with a biaxial plus a [010]
uniaxial easy axis. Note that the axis are in magnetic field units
scaled to the volume magnetization (M).}}
\end{figure}

For the model case of a purely biaxial anisotropy, this inner region
would take the form of a perfect square with corners along the easy
axis and the length of the half diagonal given by $\varepsilon/M$,
the domain wall nucleation/propagation energy scales to the volume
magnetization (Fig.~\ref{squares}a). The inclusion of a uniaxial
anisotropy bisecting two of the biaxial easy axes moves the
resulting easy axes towards the direction of the uniaxial anisotropy
\cite{Goennenwein} and elongates the square into a rectangle as
schematically depicted in Fig.~\ref{squares}b. The strength of the
uniaxial anisotropy constant in the [110] direction $K_{110}$
relative to the biaxial anisotropy constant $K_{biax}$ can be
extracted from the angle $\delta$, as defined in
Fig.~\ref{squares}b, by which the angle between two easy-axes is
modified. The relationship is given by \cite{NJPfingerprints}:

\begin{center}
{\parbox{15cm}{
\begin{equation}
\label{eqn:delta}
\delta=\arcsin\left(\frac{K_{uni[110]}}{K_{biax}}\right)
\end{equation}
}}
\end{center}

In practice, because the mixing of the anisotropy terms leads to a
rectangle with open corners, it is often more convenient to work
with the aspect ratio of the width (W) to the length (L) of the
rectangle, instead of the angle $\delta$, which is related to the
anisotropy terms as:

\begin{center}
{\parbox{15cm}{
\begin{equation}
\label{eqn:k}
\frac{K_{uni[110]}}{K_{biax}}=\cos\left(2\arctan\left(\frac{W}{L}\right)\right)
\end{equation}
}}
\end{center}

If a uniaxial anisotropy is instead added parallel to one of the
biaxial easy axes, an asymmetry arises in the energy required to
switch between the two biaxial easy axes. Essentially, the energy
required to switch towards the easier of the two biaxial easy axis
is less than that to switch towards the second biaxial. The inner
pattern is then comprised of parts of an inner and and outer square,
and the difference in the length of their half diagonal is a measure
of $K_{010}$ (Fig.~\ref{squares}c), where $K_{010}$ is the [010]
anisotropy constant. Because of deformation of the fingerprint near
the corners of the rectangle, which results from mixing of the
anisotropy terms, it is often easier to identify the presence of an
[010] uniaxial easy axis by looking at the spacing between the sides
of the squares (or rectangles in the case that a [110] uniaxial term
is also present), as indicated by the yellow line in
Fig.~\ref{squares}c, which of course has a length equal to
$\sqrt{2}K_{010}$

\section{Characteristic of a wafer}\label{sec:wafer}

\begin{figure}[h!] \centering
\includegraphics[width=0.85\textwidth]{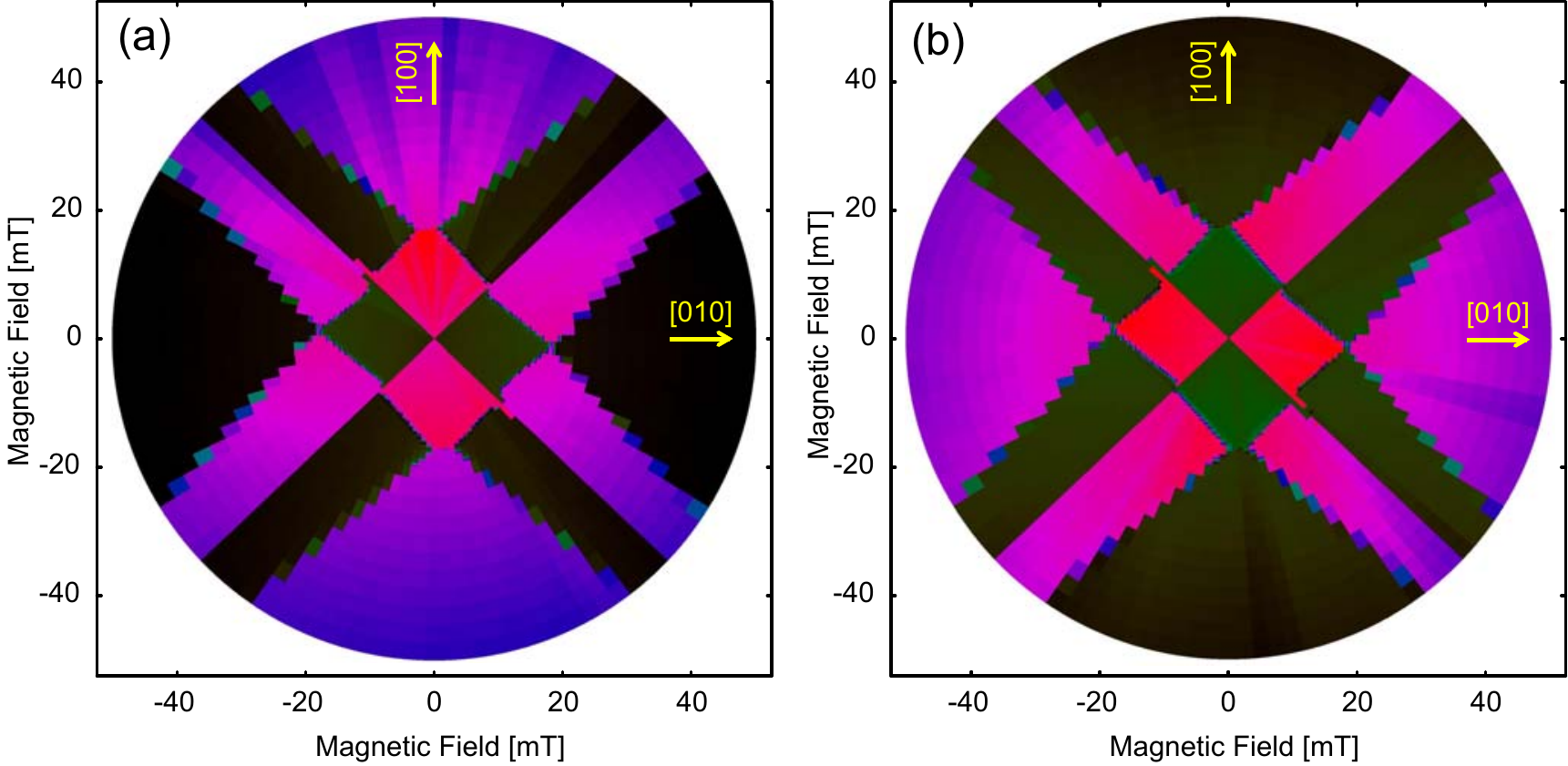}
\caption{\label{uniform} \textit{ Resistance polar plots taken on
two different locations of the same (Ga,Mn)As layer.} }
\end{figure}

We have previously shown \cite{NJPfingerprints} that the fingerprint
technique can be used to characterize the properties of a given
wafer, and for macroscopic sized devices the fingerprint is a
signature of the underlying material. As an example of this, we
present in Fig. \ref{uniform} the fingerprints for two Hall bars
patterned from different locations on the same (Ga,Mn)As wafer, and
oriented orthogonal to each other. An inspection of both
fingerprints shows that the pattern is identical, as would be
expected from an homogenous wafer. The colors are inverted because
of the $90^{\circ}$ difference in current orientation. Both
fingerprints yield the values of 1 mT for $K_{010}/M$, 18 \% for
$K_{110}/K_{biax}$ and 18 mT for $\epsilon/M$.

\begin{figure}[h!] \centering
\includegraphics[width=0.9\textwidth]{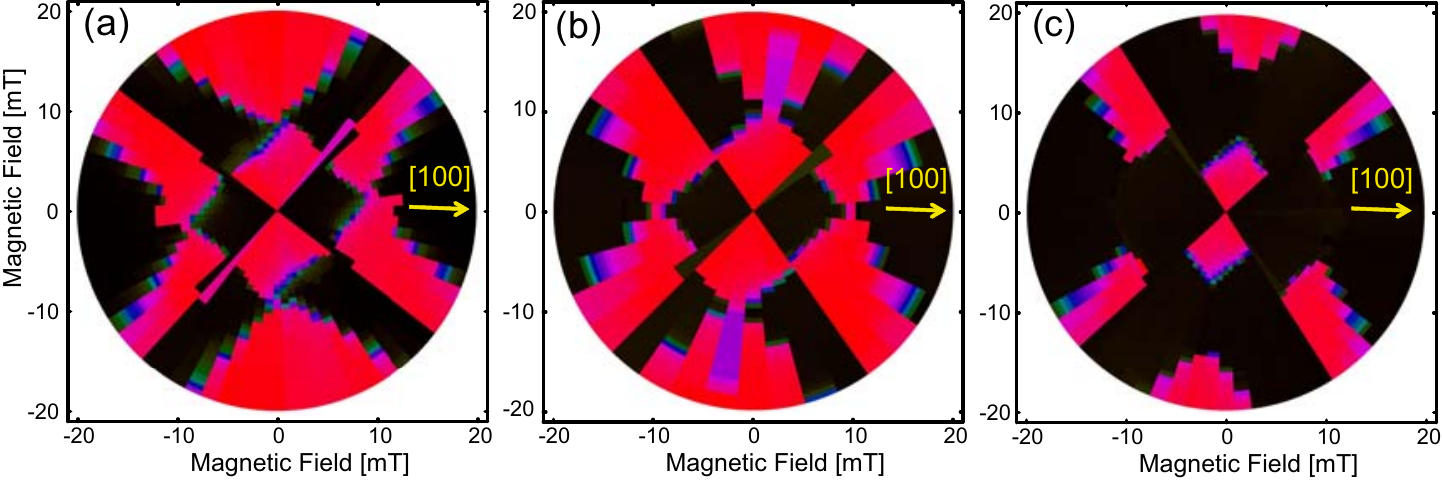}
\caption{\label{edge} \textit{Resistance polar plots taken a) near
the center of a (Ga,Ma)As wafer, b) about 4 mm from the edge of the
wafer, and c) about 1 mm from the edge.} }
\end{figure}

Next, we demonstrate how this technique can be used also as quality
control process. Figure \ref{edge} shows 3 fingerprints from three
pieces of the same (Ga,Mn)As layer, taken near the center of the 2
inch wafer, and 4 and 1 mm from the edge. Because of the geometry of
the W\"{u}rzburg MBE chamber, and given that the substrate is
rotated during growth to enhance radial homogeneity, the uniformity
of the sample is nearly perfect near the center and any fingerprint
taken in that region is identical to that of Fig.~\ref{edge}a. From
the figure we see that the central part of the sample has rather
typical values of 16 \% for $K_{110}/K_{biax}$, 0.7 mT for
$K_{010}/M$, and 9.2 mT for $\epsilon/M$. Because of non-linearities
in the molecular beam profile, stoichiometric deviations in the
epilayer become significant near the edge of the wafer. The
outermost 5 mm of samples are thus significantly less uniform. This
area is normally discarded, and certainly not used for device
studies. The fingerprint in Fig. \ref{edge}c, taken on a piece 1 mm
from the edge very clearly shows why. It presents a fingerprint
pattern very different from the homogenous center, with an enormous
discontinuity in the edges of the squares corresponding to a very
large value of $K_{010}/M=3.8$ mT for the [010] uniaxial anisotropy
term. This is well outside the range of what is found on the
homogeneous part of any (Ga,Mn)As. The deformation is sufficient
that it is impossible to reliably extract values $K_{110}$ or
$\epsilon$. The fingerprint of Fig.~\ref{edge}b, on a piece 4 mm
from the edge, is just outside the region that is normally
considered usable. It has approximately the same value of $\epsilon$
and $K_{010}$ as the central part and is only slightly deformed with
a smaller $K_{110}/K_{biax}$ of 12\%. These numbers are still within
the typical range for (Ga,Mn)As, but show a change in layer
properties as one approaches the edge.

\section{Comparison of wafers from multiple sources}\label{sec:compare}

In order to confirm that the coexistence of both the [010] and [110]
uniaxial anisotropy terms are not a particularity of (Ga,Mn)As grown
in a certain MBE chamber or under particular conditions, but are
indeed ubiquitous to the material, we now present the results of
measurements performed on samples patterned from layers grown in
various laboratories and thus under varied growth conditions.

\begin{figure}[h!] \centering
\includegraphics[width=0.88\textwidth]{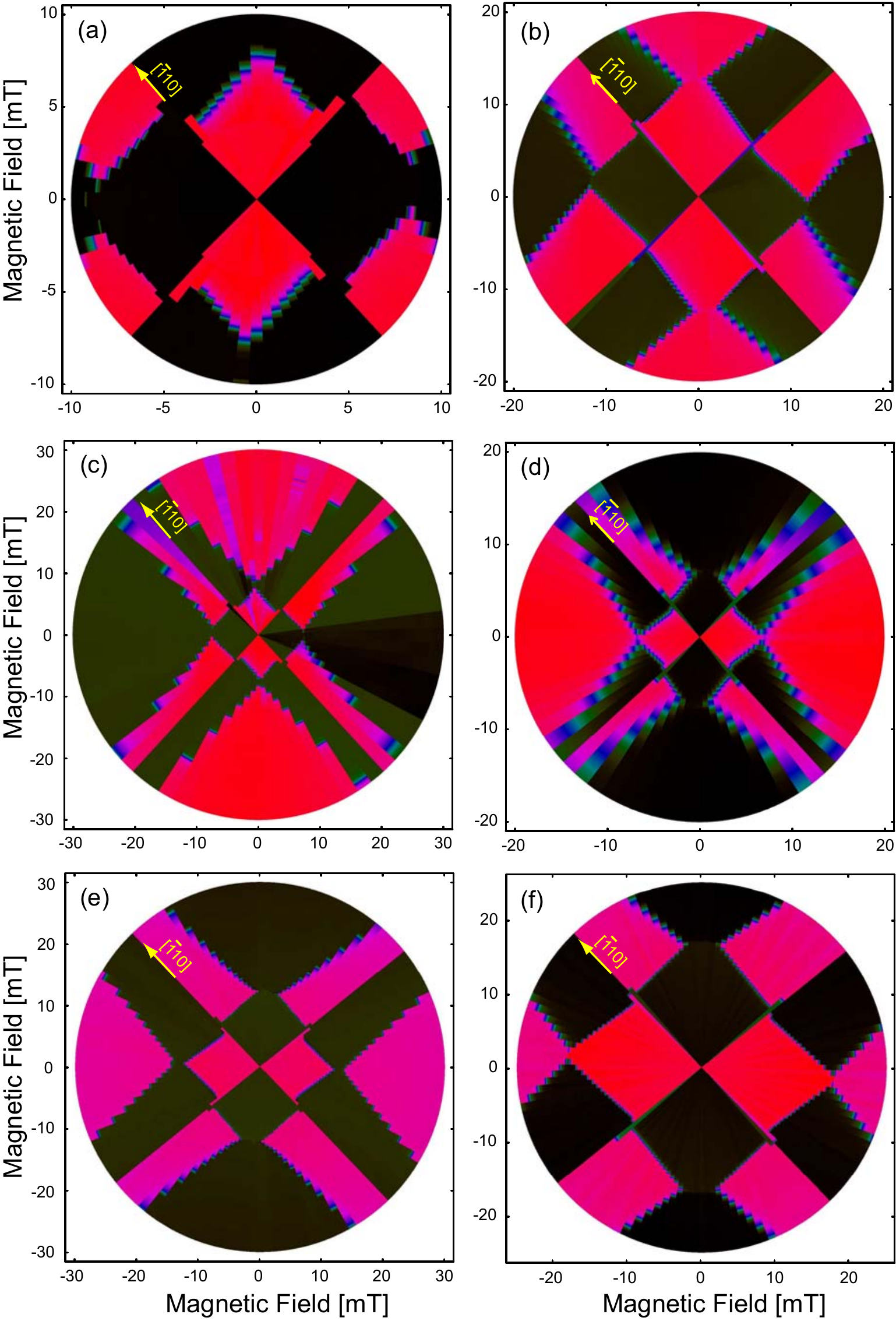}
\caption{\label{ww} \textit{Fingerprints from (Ga,Mn)As layers grown
in various laboratories. a) and b) are layers grown in W\"{u}rzburg
with strong [010] and [110] easy axis, respectively. The other
fingerprints are from layers grown at c) IMEC, d) Nottingham, e)
Tohoku, and f) Notre Dame.} }
\end{figure}

Figure \ref{uniform} showed fingerprints from a fairly typical layer
grown in W\"{u}rzburg, albeit one with a relatively large domain
wall nucleation propagation energy. To illustrate the typical spread
that can be expected, we present in Fig.~\ref{ww} two additional
W\"{u}rzburg layers with rather pronounced [010] (Fig.~\ref{ww}a) or
[110] (Fig.~\ref{ww}b) components. In parts c-f of the figure we
compare these to fingerprints on layers grown at IMEC, Nottingham,
Tohoku, and Notre Dame. Values of the various parameters extracted
from all these layers are given in Table 1. The figure illustrates
that not only the amplitude, but also the sign of the two uniaxial
components can vary between samples. For the [110] uniaxial, this
change in sign can be seen by a $90^{\circ}$ rotation of the long
axis of the rectangle, whereas the sign of the [010] is determined
by whether the quarter of the rectangle with its primary diagonal
along [010] is larger or smaller than that with the diagonal along
[100]. Note that the sign of the color scale (determining which
regions or red and which are black) is determined by the direction
of the current flow during the measurement, and is irrelevant to the
current investigation.

\begin{table}
\begin{center}
\begin{tabular}{l|r|r|r|}
 & $\epsilon/M (mT)$ & $K_{110}/K_{biax} (\%)$ & $K_{010}/M (mT)$ \\
\hline
W\"{u}. from Fig. 4 & 18 & 18 & 1.0 \\
\hline
W\"{u}. with large [010] & 8.5 & 7 & 1.4 \\
\hline
W\"{u}. with large [110] & 12 & 21 & 0.7 \\
\hline
IMEC & 7.8 & 11 & 0.7 \\
\hline
Nottingham & 7.1 & 9 & 0.65 \\
\hline
Tohoku & 12 & 4 & 1.25 \\
\hline
Notre-Dame & 16 & 9 & 0.75 \\
\hline
\end{tabular}
\caption{Characterization parameters extracted from the anisotropy
fingerprints on various layers.}
\end{center}
\end{table}

As is clear from the table, all samples show a significant
contribution of both a [110] and [010] uniaxial anisotropy
component. The values of the parameters that can be extracted from
the fingerprints show variance from sample to sample, and typically
fall in the range of some 7 to 18 mT for $\epsilon/M$, 0.6 to 1.5 mT
for $K_{010}/M$, 4 to 20\% for the ratio of $K_{110}/K_{biax}$. Note
that while the fingerprint technique cannot be used to reliably
extract exact values for $K_{biax}$, the shape of the curve as the
magnetization rotates away from the easy axis towards the external
magnetic field at higher fields can be used to estimate the strength
of $K_{biax}/M$. All samples investigated showed a value of
approximately 100 mT for this parameter which means that the values
of $K_{110}/K_{biax}$ quoted in percent in the table are also
estimates of $K_{110}/M$ in mT.

While the table clearly shows significant variation from sample to
sample, it nevertheless allows the extraction of useful rules of
thumb for relative amplitude of the various terms. As a general
statement, the ratio of $K_{biax}:K_{110}:K_{010}$ is of order
$100:10:1$, and the domain wall nucleation/propagation energy is of
the order of 10\% of the biaxial anisotropy constant.

The range of values for $K_{010}/M$ and $\epsilon/M$ seen in the
samples discussed in this study is a fair representation of
(Ga,Mn)As in general. The span of values for the $K_{110}/K_{biax}$
ratio, which is already in the table larger than the other
parameters, is however only a reflection of the subset of samples
that we investigated. In general, this ratio can easily be tuned
over a much larger range, for example as a function of hole
concentration \cite{Sawicki_InPlane} or of temperature
\cite{NJPfingerprints}. No systematic distinction is observed
between samples from various source.

\section{Conclusions}\label{sec:conclud}

In conclusion, we have used the anisotropy fingerprint technique to
analyze the magnetic anisotropy properties of multiple (Ga,Mn)As
layers. We have shown that this technique is a reliable means of
characterizing a given layer, and that it can be used as a quality
control check of the growth. Moreover, we have examined pieces of
(Ga,Mn)As grown in various laboratories around the world, and found
that all samples exhibit three magnetic anisotropy components: A
biaxial anisotropy along ([100] and [010]), a uniaxial along [110]
(or $[\bar{1}10]$) and a second uniaxial along either [100] (or
[010]), showing that the existence of all three terms is an inherent
property of (Ga,Mn)As, and that it is only the relative strength of
the terms which varies from sample to sample. As a rough rule of
thumb, the ration of the biaxial anisotropy, the [110] uniaxial and
the [010] uniaxial is of order $100:10:1$.

\section*{Acknowledgements}
The authors wish to thank S. H\"{u}mpfner and V. Hock for sample
preparation, and acknowledge the financial support from the EU
(NANOSPIN FP6-IST-015728) and the National Science Foundation (US)
Grant DMR06-03752.

\section*{References}

\bibliographystyle{prsty}
\bibliography{Spintronics}

\begin{thebibliography}{10}

\bibitem{PhysicaE}
F. Matsukura, M. Sawicki, T. Dietl, D. Chiba, and H. Ohno, Physica E {\bf 21},
  1032  (2004).

\bibitem{Liu2003}
X. Liu, Y. Sasaki, and J.~K. Furdyna, Phys. Rev. B {\bf 67},  205204  (2003).

\bibitem{SawickiAnis}
M. Sawicki, F. Matsukura, A. Idziaszek, T. Dietl, G.~M. Schott, C. Ruester, C.
  Gould, G. Karczewski, G. Schmidt, and L.~W. Molenkamp, Phys. Rev. B {\bf 70},
   245325  (2004).

\bibitem{Sawicki_InPlane}
M. Sawicki, K.~Y. Wang, K. Edmonds, R. Campion, C. Staddon, N. Farley, C.
  Foxon, E. Papis, E. Kamiska, A. Piotrowska, T. Dietl, and B. Gallagher, Phys.
  Rev. B {\bf 71},  R121302  (2005).

\bibitem{Roukes}
H.~X. Tang, R.~K. Kawakami, D.~D. Awschalom, and M.~L. Roukes, Phys. Rev. Lett.
  {\bf 90},  107201  (2003).

\bibitem{TAMR1}
C. Gould, C. R\"uster, T. Jungwirth, E. Girgis, G.~M. Schott, R. Giraud, K.
  Brunner, G. Schmidt, and L.~W. Molenkamp, Phys. Rev. Lett. {\bf 93},  117203
  (2004).

\bibitem{FurdynaGrowth}
J. Furdyna, X. Liu, N. Lim, Y. Sasaki, T. Wojtowicz, and I. Kuryliszyn, Journ.
  Korean Phys. Soc {\bf S579},  2003  (42).

\bibitem{CampionGrowth}
R. Campion, K. Edmonds, L. Zhao, K. Wang, C. Foxon, B. Gallagher, and C.
  Staddon, Journ. Cryst. Growth {\bf 247},  20  (2003).

\bibitem{Giselagrowth}
G. Schott, G. Schmidt, G. Karczewski, L.W.Molenkamp, R. Jakiela, and A. Barcz,
  Appl. Phys. Lett. {\bf 82},  4678  (2003).

\bibitem{Fingerprints}
K. Pappert, S. H\"{u}mpfner, J. Wenisch, K. Brunner, C. Gould, G. Schmidt, and
  L.~W. Molenkamp, Appl. Phys. Lett. {\bf 90},  062109  (2007).

\bibitem{NJPfingerprints}
K. Pappert, C. Gould, M. Sawicki, J. Wenisch, K. Brunner, G. Schmidt, and L.~W.
  Molenkamp, New Journ. Phys. {\bf 9},  354  (2007).

\bibitem{Baxter}
D.~V. Baxter, D. Ruzmetov, J. Scherschligt, Y. Sasaki, X. Liu, J.~K. Furdyna,
  and C.~H. Mielke, Phys. Rev. B {\bf 65},  212407  (2002).

\bibitem{Jan1957}
J.~P. Jan, {\em Solid State Physics} ((Eds: F. Seitz, D. Turnbull), Academic
  Press Inc., New York, 1957), pp.\ Vol. 5, p.1.

\bibitem{McGuire}
T. McGuire and R. Potter, IEEE Trans. Magn. {\bf MAG-11},  1018  (1975).

\bibitem{Rushforth}
A.~W. Rushforth, K. Vyborny, C.~S. King, K.~W. Edmonds, R.~P. Campion, C.~T.
  Foxon, J. Wunderlich, A.~C. Irvine, P. Vasek, V. Novak, K. Olejník, J.
  Sinova, T. Jungwirth, and B.~L. Gallagher, Phys. Rev. Lett. {\bf 99},  147207
   (2007).

\bibitem{Welp03PRL}
U. Welp, V. Vlasko-Vlasov, X. Liu, J. Furdyna, and T. Wojtowicz, Phys. Rev.
  Lett. {\bf 90},  167206  (2003).

\bibitem{Goennenwein}
S.~T.~B. Goennenwein, S. Russo, A.~F. Morpurgo, T.~M. Klappwijk, W. van Roy,
  and J. de~Boeck, Phys. Rev. B {\bf 71},  193306  (2005).

\end{thebibliography}

\end{document}